\documentclass[12pt]{article}  
\begin{document}

\newcommand{\sheptitle}
{Discriminating neutrino mass models  using  Type II seesaw formula}
\newcommand{\shepauthor}
{N.Nimai Singh$^{a,c,}${\footnote{Regular Associate of ICTP,\\ {\it{e-mail address:}} nsingh@ictp.trieste.it; nimai03@yahoo.com}},
Mahadev Patgiri$^{b}$ and Mrinal Kumar Das$^{c}$ }

\newcommand{\shepaddress}
{${^a}$International Centre for Theoretical Physics, Strada Costiera 11,
\\ 31014 Trieste, Italy\\
${^b}$Department of Physics, Cotton College, Guwahati-781001, India\\
${^c}$ Department of Physics, Gauhati University, Guwahati-781014, India}
\newcommand{\shepabstract}
{In this paper we propose a kind of  natural selection which can discriminate the 
 three possible neutrino mass models, namely 
the degenerate, inverted hierarchical and normal hierachical models, using the framework of Type II seesaw formula. We arrive at a conclusion 
that the inverted hierarchical model appears to be most favourable whereas the normal hierarchical model follows next to it. The degenerate 
model is  found to be most unfavourable. The neutrino mass matrices which are  obtained using the usual canonical seesaw formula (Type I),
 and  which also  give 
almost good predictions of neutrino masses and mixings consistent with the latest neutrino oscillation data,
 are re-examined in the light of non-canonical seesaw formula (Type II). 
We then estimate a parameter $\gamma$ which represents the minimum degree of suppression  of the extra term arising from the left-handed Higgs 
triplet, so as to restore the good predictions 
on neutrino masses and mixings already had in Type I seesaw model.} 

\begin{titlepage}
\begin{flushright}
hep-ph/0406075
\end{flushright}
\begin{center}
{\large{\bf\sheptitle}}
\bigskip\\
\shepauthor
\\
\mbox{}\\
{\it\shepaddress}\\
\vspace{.5in}
{\bf Abstract}
\bigskip
\end{center}
\setcounter{page}{0}
\shepabstract
\end{titlepage}
\section{Introduction}
Recent neutrino oscillation experiments[1] which  provide important informations on the nature of
neutrino masses and mixings, have strengthened our understanding of neutrino oscillation[2,3]. 
However  we are still far from a complete understanding of neutrino physics. 
One of them is the pattern of the neutrino mass eigenvalues, though some reactor experiments are trying to understand it.
 For future reference we summarise here[1] the most recent results of the three-flavour neutrino 
oscillation parameters from global data including solar[4], atmospheric[5], reactor (KamLAND[6] and CHOOZ[7])
 and accelerator (K2K [8]):\\

\begin{tabular}{lll} \hline
parameter & best fit &  3$\sigma$ level \\ \hline
$\bigtriangleup m^2_{21}[10^{-5}eV^2]$ & $6.9$ & $5.4-9.4$ \\ 
$\bigtriangleup m^2_{23}[10^{-3}eV^2]$ & $2.3$ & $1.1-3.4$ \\
$\sin^2\theta_{12}$                    & $0.3$ & $0.23-0.39$ \\ 
 $\sin^2\theta_{23}$                    & $0.52$ & $0.32-0.70$ \\ 
 $\sin^2\theta_{13}$                    & $0.005$ & $\leq 0.061$ \\ \hline \\
\end{tabular}
\\
As  far as  the LSND result[9] is concerned, it is finding difficulty to reconcile with the rest of the global data, and 
 a confirmation of the LSND signal by the MiniBooNE experiment[10] would be  very desirable.
 There are also some  complementary information from other sources. The recent analysis 
of the WMAP collaboration[11] gives the bound 
$\sum_i|m_i|<0.69$eV (at $95\% $ C.L.)  (more conservative analysis[12] gives 
$\sum_i|m_i|<1.01$eV). The bound from the $0\nu\beta\beta$-decay experiment is
$|m_{ee}|<0.2$eV (a more conservative analysis gives 
$|m_{ee}|<(0.3 - 0.5)$eV)[13,14]. However the value of the $|m_{ee}|$ from the recent claim[15] for
 the discovery of the $0\nu\beta\beta$ process  at $4.2\sigma$ level, is 
$|m_{ee}|\sim (0.2-0.6)$eV   (more conservative estimate involving nuclear mass is  $|m_{ee}|\sim (0.1-0.9)$eV).\\

 Since the above data on solar and atmospheric neutrino oscillation experiments,  gives only the mass square differences, we  usually  have  
three models\footnote{In order to avoid possible confusion in nomenclature, 
Types of neutrino mass models are denoted inside the square bracket whereas Types of seesaw formula are expressed 
without square bracket.}    of neutrino mass levels[16]: \\
{\bf Degenerate (Type [I])}: $m_{1}\simeq m_{2}\simeq m_{3}$.,\\
{\bf Inverted hierarchical (Type [II])}: $m_{1}\simeq m_{2}>>m_{3}$ with $\bigtriangleup m^{2}_{23}=m^{2}_{3}-m^{2}_{2}<0$
 and $m_{1,2}\simeq \sqrt{\bigtriangleup m^{2}_{23}}\simeq (0.03-0.07)$eV; and\\
{\bf Normal hierarchical (Type [III])}: $m_{1}<<m_{2}<<m_{3}$, and 
$\bigtriangleup m^{2}_{23}=m^{2}_{3}-m^{2}_{2}>0$; and
$m_{3}\simeq \sqrt{\bigtriangleup m^{2}_{23}}\simeq (0.03-0.07)$ eV, and $m_{2}\simeq 0.008$eV. \\
 ( Appendix A  presents  a  list  of the 
zeroth-order left-handed Majorana mass matrices which can explain the above three patterns of neutrino masses).  
 The result of $0\nu\beta\beta$ decay experiment, if confirmed, would be able to rule out Type [II] and Type [III] neutrino mass models straight,
 and points to Type [I] or to models with more than three neutrinos[3].
Again, the WMAP limit (at least for three degenerate neutrinos), $|m|<0.23$eV also
 would rule out Type [I] neutrino model, or at least it could lower the parameter space for the degenerate model[3].
 It also gives further constraint on $|m_{ee}|$. However  a final choice among these three models is a difficult task.
At the moment we are in a very confusing state. The present paper is a modest attempt from a 
theoretical point of view to discriminate the neutrino mass models using the Type II seesaw formula 
(non-canonical seesaw formula) for neutrino masses.

The paper is organised as follows. In section 2, we outline the main points of the Type II seesaw formula and a criteria
 for a natural selection which helps to  discriminate the neutrino mass models. We carry out numerical computations in section 3 and present 
our main results. Section 4 concludes with a summary and discussions.    

\section{Type II see-saw formula and neutrino mass matrix}
The canonical  seesaw mechanism (generally known as Type I seesaw formula)[17] is the  simplest and most appealing mechanism
for generating small neutrino masses and lepton mixings. There is also  another type of seesaw formula ( known as Type II seesaw formula)[18] 
where  a left-handed Higgs triplet $\Delta_{L}$ picks up a vacuum expectation value (vev)  in the left-right symmetric GUT models 
such as SO(10). This is expressible as 
\begin{equation}
m_{LL}=m_{LL}^{II} + m_{LL}^I.
\end{equation}
 where the usual  Type I  seesaw formula  is given by the expression,
\begin{equation}
m_{LL}^I=-m_{LR}M_{RR}^{-1}m_{LR}^{T}
\end{equation}
Here  $m_{LR}$ is the Dirac neutrino mass matrix in the left-right convention and the right-handed Majorana neutrino mass matrix 
$M_{RR}=v_{R}f_{R}$ with $v_{R}$ being the vacuum expectation value (vev) of the  Higgs fields imparting mass to the
 right-hand neutrinos and $f_{R}$ is the Yukawa coupling matrix. 
The second term $m_{LL}^{II}$ is due to the $SU(2)_{L}$ Higgs triplet, which can arise, for instance, in a large class of $SO(10)$
models in which the $B-L$ symmetry is broken by a $126$ Higgs field[19,20]. In the usual left-right symmetric theories,  $m_{LL}^{II}$
 and $M_{RR}$ are proportional to the vacuum expectation values (vevs) of the electrically neutral components of 
scalar Higgs triplets, i.e., $m_{LL}^{II}=f_{L}v_{L}$ and $M_{RR}=f_{R}v_{R}$, where $v_{L,R}$ denotes the vevs and $f_{L,R}$ 
is a symmetric $3\times3$ matrix. By acquiring the vev $v_{R}$, breaking of $SU(2)_{L}\times SU(2)_{R}\times U(1)_{B-L}$ to
 $SU(2)_{L}\times U(1)_{Y}$ is achieved. The left-right symmetry demands the presence of both $m_{LL}^{II}$ and $M_{RR}$, and 
in addition, it holds $f_{R}=f_{L}=f$. The induced vev for the left-handed triplet $v_{L}$ is given by  $v_{L}=\gamma M^{2}_{W}/v_{R}$,
 where the weak scale $M_{W}\sim 82$GeV such that $|v_{L}|<<M_{W}<<|v_{R}|$[21]\footnote{In some papers[19] $v_{u}\sim 250$GeV is taken
in place of $M_{W}$. We prefer here to take $\sim 82$GeV as it is nearer to our input value of either $m_{t}=82.43$GeV or $m_{\tau}\tan\beta=1.3\times 40$GeV
in the text. In this way  both the terms of the Type II seesaw formula, have almost same value of  weak scale. However taking different values does not alter 
the conclusion of our analysis}. In general $\gamma$ is a function of various couplings,
 and without fine tuning $\gamma$ is expected to be of the order unity. Type II seesaw formula in Eq.(1)
can now be expressed as
\begin{equation}
m_{LL}=\gamma (M_{W}/v_{R})^{2}M_{RR}-m_{LR}M^{-1}_{RR}m^{T}_{LR}
\end{equation}  

In the light of the above Type II seesaw formula Eq.(3), the neutrino mass matrices, $m_{LL}$ in literature, 
are constructed in view of the following three assumptions:
(a) $m_{LL}^{II}$ is dominant over $m_{LL}^{I}$, and 
(b) both  terms are contributing with comparable amounts,
(c) $m_{LL}^{I}$ is dominant over $m_{LL}^{II}$. In recent times case(a) has gathered  
momentum because in certain $SO(10)$ models, large atmospheric neutrino mixing and $b-\tau$ unification
are the  natural outcomes of this dominance[20,22,23]. In some models this leads to degenerate model [24] which imparts bimaximal mixings, as well as 
extra contribution to leptogenesis[24,25,26]. However all these cases are not completely free from certain  assummptions as well as  ambiguities. 

It  can be stressed that these two terms $m_{LL}^{I}$ and $m_{LL}^{II}$ in Eq.(1), are not completely independent.
 The term $m_{LL}^{II}$ is heavily
 constrained through $v_{R}$ as seen in Eq.(3).  Usually the value of $v_{R}$ is  fixed through the definition $M_{RR}=v_{R}f$ 
present in the  canonical  term $m_{LL}^{I}$.
 There is  no ambiguity in the definition of $v_{R}$ with the first term, and  it also does not affect $m_{LL}^{I}$ as long as $M_{RR}$ 
is taken as a whole in the expression.
However it severly affects the second term $m_{LL}^{II}$ where $v_{R}$ is entered alone, and different choices of $v_{R}$ in $M_{RR}$  
would lead to different values of $m_{LL}^{II}$.
This ambiguity is seen in the literature where different choices of $v_{R}$ are made according to the convenience[19,21,22,25,28]. However, in the 
present paper we shall always take $v_{R}$ as the heaviest right-handed Majorana neutrino mass eigenvalue obtained after the diagonalisation of the mass 
matrix $M_{RR}$.  
Once we adopt this convention, 
there is  little freedom for the second term $m_{LL}^{II}$ in Eq.(3) to have arbitrary   
value of  $v_{R}$. We also assume that the $SU(2)_{R}$ gauge symmetry breaking scale $v_{R}$ is the same as the scale of the  breakdown of parity [20]\footnote
{See Ref.[37] for further discussion on the choice of $v_{R}$}.

The  present work is carried out in the line of case (b) and (c) discussed above, but the choice of which term is dominant over other, is not arbitrary any more.
 We carry out a complete analysis of the three models of
neutrino mass matrices (See Appendix B for the expressions of $M_{RR}$ and $m_{LL}$) where the (already acquired) good predictions of
 neutrino masses and mixings in the canonical 
term $m_{LL}^{I}$ is subsequently spoiled 
by the presence of second term $m_{LL}^{II}$ when $\gamma=1$ in $m_{LL}$.  We  make a search programme for finding the  
values of the minimum departure of $\gamma$ from of the canonical value of one, in which  the good predictions of neutrino
 masses and mixing parameters can be restored in $m_{LL}$. We  propose here a  bold hypothesis which acts as a sort of `` natural selection'' for the survival 
of the neutrino mass models
which enjoy the least value of deviation of $\gamma$ from unity. In other words, the value of $\gamma$ is enough just to 
suppress the 
perturbation effect arising from Type II seesaw formula. Nearer the value of $\gamma$ to one, better the chance for the survival 
of the model in question. Thus the parameter $\gamma$ is an important parameter  for the proposed natural 
selection of the neutrino mass models.

The above criteria for natural selection  imposes certain constraints on the neutrino mass models which one can obtain 
in the following way, at least for the heaviest neutrino mass eigenvalue (without considering mixings).
If the neutrino masses are solely determined from the second term of Eq.(3), then the first term must be less than 
the certain order which is dictated by the particular pattern of neutrino mass spectrum. 
In this view, the largest contribution of neutrino mass from the first term must be 
less than about $0.05$eV for both normal hierarchical and inverted hierarchical models;  and
 about $0.5$eV for degenerate model as the data suggests[1].
Thus we have the bound for the natural selection:\\
\begin{equation}
m_{LL}^{I}>v_{L}f 
\end{equation}
 Denoting the heaviest right-handed neutrino mass as $v_{R}$ and
 taking $M_{W}\sim 82$GeV[21] in the expression of $v_{L}$, the following  lower bounds on $v_{R}$  for the natural selection are obtained:\\

{\bf For normal hierchical and inverted hierarchical model:}
\begin{equation}
v_{R}>\gamma 1.345\times 10^{14} GeV
\end{equation}

{\bf For degenerate model:}
\begin{equation}
v_{R}> \gamma 1.345\times 10^{13} GeV
\end{equation}
The above bounds just indicate the approximate measure of the degree of natural selection, but a 
fuller analysis will take both the terms of the Type II seesaw formula in the $3\times3$ matrix form. This will gives all 
the three mass eigenvalues as well as mixing angles. This numerical analysis will be carried out in the next section.
It is clear from Eqs.(5) and (6) that any amount of arbitrariness in fixing the value of $v_{R}$ in $M_{RR}$ will distort the conclusion.   
\section{Numerical calculations and results}
For a full numerical analysis we refer to our earlier papers[27] where  we performed the investigations on the origin of neutrino masses and
mixings which can accomodate LMA MSW solution for solar neutrino anomaly and the solution of 
atmospheric neutrino problem within the framework of Type I seesaw formula. Normal hierarchical,
inverted hierarchical and quasi-degenerate neutrino mass models were constructed from the nonzero
textures of the right-handed Majorana mass matrix $M_{RR}$ along with diagonal form of $m_{LR}$ being 
taken as either the charged 
lepton mass matrix (case i)[28] or the up-quark mass matrix (case ii)[27]. However, a general form of the  Dirac neutrino mass matrix is given by :
 \begin{equation}
m_{LR}=\left(\begin{array}{ccc}
\lambda^{m} & 0 & 0\\
 0 & \lambda^{n} & 0 \\
 0 & 0 & 1 
\end{array}\right) m_{f},
\end{equation}
where $m_f$ corresponds to $(m_{\tau}\tan\beta)$ for $(m,n)=(6,2)$ in case of  charged lepton ({\bf case i})
 and $m_{t}$ for $(m,n)=(8,4)$ in case of  up-quarks ({\bf case ii}). The value of the parameter $\lambda$ is taken as $0.22$. Here the assumption 
 is that neutrino mass mixings 
can arise from the texture of right handed neutrino mass matrix only through the interplay of 
seesaw mechanism[29]. This can be understood from the following operation[30,31]  where $M_{RR}$ can  be 
transformed in the basis in which   $m_{LR}$ is approximately diagonal\footnote{ This is also true for any diagonal $m_{LR}$ with any arbitrary pair of $(m,n)$.
 A corresponding $M_{RR}$ can be found out in principle}. Using the diagonalisation relation, 
$m_{LR}^{diag}=U_{L}m_{LR}U_{R}^{\dagger}$, we have,  
$$m_{LL}^{I}
=-m_{LR}M_{R}^{-1}m_{LR}^{T}$$\\
$$=-U_{L}^{+}m_{LR}^{diag}U_{R}M_{R}^{-1}U_{R}^{T}m_{LR}^{diag}U_{L}^*$$\\
$$=-U_{L}^{+}m_{LR}^{diag}M_{RR}^{-1}m_{LR}^{diag}U_{L}^{*}$$\\
$$\simeq - m_{LR}^{diag}M_{RR}^{-1}m_{LR}^{diag}$$\\
where $U_{L}m_{LL}^{I}U_{L}^T\simeq m_{LL}^{I}$ by considering 
 a simple assumption, $U_{L}\simeq1$. Since the Dirac neutrino mass matrices are hierarchical in nature and the CKM mixing 
angles of the quark sector are relatively small. In such situation  $U_{L}$ slightly deviates from $1$, 
i.e., $U_{L}\simeq V_{CKM}$, and it hardly affects the numerical accuracy[30] for practical purposes. Here $M_{RR}$ is the new 
RH matrix defined in the basis of diagonal $m_{LR}$ matrix. We thus express $M_{RR}$ in the most general form as its origin is quite 
different from those of the Dirac mass matrices in an underlying grand unified theory.
As usual the neutrino mass eigenvalues and neutrino mixing matrix (MNS) are obtained through the diagonalization of  $m_{LL}$,
$$m_{LL}^{diag}=V_{\nu L}m_{LL} V_{\nu L}^{T}=Diag(m_{1},m_{2},m_{3}),$$ and the neutrino mixing angles are extracted  from
 the MNS lepton mixing matrix defined by $V_{MNS}=V_{\nu L}^{\dagger}$.
\\

{\bf Normal hierarchical model (Type [III]):}\\

We then perform a detailed numerical analysis to search for the parameter $\gamma$ which measures the perturbation effects arising from the 
Type II seesaw term. As a simplest  example, we take up the case for the 
normal hierarchical model (Type [III]) while the 
expressions for other models are relegated to Appendix B. Using 
the general expression for $m_{LR}$ given in Eq.(7) and the following texture for  $M_{RR}$ [27]:
 \begin{equation}
M_{RR}=\left(\begin{array}{ccc}
\lambda^{2m-1} & \lambda^{m+n-1} & \lambda^{m-1}\\
 \lambda^{m+n-1} & \lambda^{m+n-2} & 0 \\
 \lambda^{m-1} & 0 & 1 
\end{array}\right) v_{R},
\end{equation}
we get  the neutrino mass matrix of the Type [III] through Eq.(2),
\begin{equation}
-m_{LL}^{I}=\left(\begin{array}{ccc}
-\lambda^{4} & \lambda & \lambda^{3}\\
 \lambda & 1-\lambda & -1 \\
 \lambda^{3} & -1 & 1-\lambda^{3} 
\end{array}\right)\times 0.03 eV
\end{equation}
Here we have  fixed  the value of  $v_{R}=8.92\times 10^{13}$GeV for case (i), taking $(m,n)$ as $(6,2)$ and  the input values 
 $m_{\tau}=1.292$GeV, and $\tan\beta=40$.
The diagonalization of $M_{RR}$ gives the three corresponding RH Majorana neutrino masses 
$M_{RR}^{diag}=(-4.8555\times 10^8, 1.058\times 10^{10}, 8.92\times 10^{13})$GeV. As already stated, the mass 
matrix in Eq.(9) predicts correct neutrino mass parameters and mixing angles consistent with recent data[27]:\\

$m_{LL}^{ diag}=(0.0033552,0.0073575,0.057012)$eV,  leading to \\
$\bigtriangleup m^{2}_{21}=4.29\times 10^{-5}$eV, $\bigtriangleup m^{2}_{23}=3.20\times 10^{-3}$eV,
 $\alpha=\bigtriangleup m^{2}_{21}/\bigtriangleup m^{2}_{23}=0.0134$. \\
$\sin\theta_{12}=0.5838$, $\sin\theta_{23}=0.6564$, $\sin\theta_{13}=0.074.$\\
\\

In the next step we take the additional contribution of the second term $m_{LL}^{II}=\gamma (M_{W}/v_{R})^2 M_{RR}$ in Type II seesaw formula
 in Eq.(3). When $\gamma=1$,  all the good predictions of neutrino masses and mixings already had in $m_{LL}^{I}$, are then spoiled. The value of $\gamma$
for the ``least deviation from canonical value of one'', which could restore the good predictions in $m_{LL}$, is again obtained through 
a search programme. The predictions are:  $\gamma\simeq 0.1$,  
$m_{LL}^{ diag}=(-0.0021353,0.0095481,-0.0534155)$eV,  leading to \\
$\bigtriangleup m^{2}_{21}=8.66\times 10^{-5}$eV, $\bigtriangleup m^{2}_{23}=2.76\times 10^{-3}$eV,
 $\alpha=\bigtriangleup m^{2}_{21}/\bigtriangleup m^{2}_{23}=0.0314$. \\

The corresponding MNS mixing matrix which diagonalizes $m_{LL}$ is obtained as
 \begin{equation}
V_{MNS}=\left(\begin{array}{ccc}
  0.879342 & 0.468733 & -0.083946\\
 0.275275 & -0.644212 & -0.713593\\
 0.388564 & -0.604384 & 0.695513 
\end{array}\right)
\end{equation}
leading to 
$\sin^{2}2\theta_{12}=0.6796$, $\tan^{2}\theta_{12}=0.284<1$, $\sin^{2}2\theta_{23}=0.98531$, $\sin\theta_{13}=0.084.$\\
Here we have given solar mixing angle in term of $tan^{2}\theta_{12}$ to check whether it falls in the ``light side'',  $\tan^{2}\theta_{12}<1$, 
for the usual sign convention $\bigtriangleup m^{2}_{21}=m_{2}^{2}-m_{1}^{2}>0, [32,33]$.
For the case (ii) when $(m,n)=(8,4)$ in Eq.(7), we take the input value $m_{t}=82.43$GeV at the high scale. 
We have again the final predictions from $m_{LL}$:
 $\gamma\simeq0.1$, and 
$M_{RR}^{diag}=(-2.891\times 10^6, 6.299\times 10^7, 2.267\times 10^{14})$GeV,\\
$\bigtriangleup m^{2}_{21}=5.81\times 10^{-5}$eV, $\bigtriangleup m^{2}_{23}=3.02\times 10^{-3}$eV,
 $\alpha=\bigtriangleup m^{2}_{21}/\bigtriangleup m^{2}_{23}=0.0192$, \\
$\sin^{2}2\theta_{12}=0.8233$, $\tan^{2}\theta_{12}=0.42<1$, $\sin^{2}2\theta_{23}=0.9862$, $\sin\theta_{13}=0.077$.\\

We also calculate the mass parameter $|m_{ee}|$ measured in the $0\nu\beta\beta$ decay experiment using the expression[3]
\begin{equation}
|m_{ee}|=|(1-\sin^{2}\theta_{13})(m_{1}\cos^{2}\theta_{12}+m_{2}\sin^{2}\theta_{12})+m_{3}e^{2i\phi}\sin^{2}\theta_{13}|
\end{equation}
For degenerate (Type [I]) and inverted hierarchy (Type [II]) model, it reduces to (for $\sin^{2}\theta_{13}\simeq 0$)
\begin{equation}
|m_{ee}|\sim|m|(\cos^{2}\theta_{12}\pm\sin^{2}\theta_{12})
\end{equation}
In case of normal hierarchy (Type [III]), we have 
\begin{equation}
|m_{ee}|\sim|m_{2}\sin^{2}\theta_{12}\pm m_{3}\sin^{2}\theta_{13}|
\end{equation}
For the example discussed above (normal hierarchy for case (i)), we obtain  $|m_{ee}|=0.0017$.

In {\bf Appendix B}  we list all the $m_{LL}^{I}$ along with the  corresponding $M_{RR}$ textures for Degenerate (Type [I (A,B,C)]) and  Inverted hierarchy
 (Type [II(A,B)])[27]. We repeat the same procedure described above for all these cases and find out the corresponding values of $\gamma$.\\

We present here  the main results of the analysis.
We calculate RH neutrino masses in Table-1 for both cases (i) and (ii). The heaviest RH Majorana eigenvalue 
is taken as $v_{R}$ scale for calculation of $m_{LL}^{II}$. It is interesting to see in Table-1 that only Type [II(A,B)] satisfie the bounds 
given in Eqs.(5) and (6) when $\gamma=1$. This roughly implies that Inverted hierarchical model is the best choice for natural selection though a fuller analysis 
needs  the matrix form when all terms are present. 
 Our main results on neutrino masses and mixings are presented in Table-2 and Table-3.
One particular important parameter is the predicted values of $\gamma$. Table-4 presents the mass parameter $|m_{ee}|$ and $\alpha$ for both cases (i) and (ii).

From Table-2 and -3, we wish to make  a  conclusion that Inverted hierarchy model (Type [II]) having $\gamma=1$ 
is the most stable under the presence of $SU(2)_{L}$ triplet term $m_{LL}^{II}$ in the Type [II] seesaw formula. On such ground we can discriminate 
other models in favour of it. Next to Inverted hierarchy is the normal hierarchy model (Type [III]) with $\gamma=0.1$. In the  present analysis 
 the degenerate model (Type [I]) is not favourable  as it predicts $\gamma\sim 10^{-4}$. Again, in Inverted hierarchy model we have
two Types - [IIA] and [IIB]. Type [IIA]   predicts 
slightly lesser  solar mixing angle and exactly zero CHOOZ angle without any fine tunning such as contribution
from charged lepton etc., compared to those of Type [IIB] where we have maximal solar mixing. The present analysis has limitation to further 
discriminate either of these two.\\

As a routine calculation we check the stability of these models under radiative corrections in MSSM for 
both neutrino mass splittings and mixing angles. For large $\tan\beta=55$ where the effect of radiative correction is relatively large, only two models namely,  
Inverted hierarchy [38] of Type [IIB] and Normal hierarchy [39] of Type [III] are found stable under radiative corrections. Following this result, the inverted 
hierarchy of Type [IIA] is less favourable than its counterpart Type [IIB].   


Table-1: The three right-handed Majorana neutrino masses for both case (i)
and case (ii) in three pattern of neutrino mass models. The $B-L$ symmetry 
breaking scale $v_{R}$ is taken as the mass of the heaviest right-handed Majorana neutrino
 in the calculation.
\begin{center}
\begin{tabular}{ccc}\hline
Type & Case(i):$|M_{RR}^{diag}|$GeV & Case(ii):$|M_{RR}^{diag}|$GeV\\ \hline
\\
\ [IA] & $6.82\times10^{9}$, $5.51\times10^{9}$, $5.04\times10^{12}$;& 
$6.34\times10^{7}$, $1.94\times10^{8}$, $8.55\times10^{12}$\\
\\
 \ [IB] & $1.50\times10^{5}$, $2.72\times10^{10}$, $1.16\times10^{13}$;& 
$5.17\times10^{2}$, $9.37\times10^{7}$, $1.7\times10^{13}$\\
\\
 \ [IC] & $1.3\times10^{5}$, $4.61\times10^{11}$, $5.06\times10^{11}$;& 
$5.17\times10^{2}$, $1.89\times10^{10}$, $8.5\times10^{10}$\\
\\
\hline
\\
 \ [IIA] & $1.19\times10^{6}$, $4.32\times10^{11}$, $4.63\times10^{17}$;& 
$4.1\times10^{3}$, $1.49\times10^{8}$, $6.8\times10^{17}$\\
\\
\ [IIB] & $2.97\times10^{8}$, $2.97\times10^{8}$,$1.16\times10^{16}$;& 
$1.74\times10^{6}$,$1.74\times10^{6}$,$2.89\times10^{16}$\\

\\
\hline
\\
 \ [III] & $4.86\times10^{8}$, $1.06\times10^{10}$, $8.92\times10^{13}$;& 
$ 2.89\times10^{6}$, $6.3\times10^{7}$, $2.27\times10^{14}$\\
\\
\hline
\end{tabular}
\end{center}

\pagebreak

Table-2:Predicted values of $\gamma$ and its corresponding  solar and atmospheric mass-squared 
differences and mixing parameters from $m_{LL}$ using the values of $v_{R}$ from Table 1 for case (i).
\begin{center}
\begin{tabular}{ccccccc}\hline
Type & $\gamma$ & $\bigtriangleup m^{2}_{21}[10^{-5}eV^{2}]$ & 
 $\bigtriangleup m^{2}_{23}[10^{-3}eV^{2}]$ & $\tan^{2}\theta_{12}$
 & $\sin^{2}_2\theta_{23}$ & $\sin\theta_{13}$      \\ \hline
\\
 \ [IA] & $10^{-4}$ & $8.56$ & $2.76$ & $1.013$ & $0.993$ & $0.03$\\
\\
  \ [IB] & $10^{-3}$ & $3.97$ & $2.48$ & $0.278$ & $1.00$ & $0.0$\\
\\
 \ [IC] & $10^{-3}$ & $3.65$ & $2.46$ & $2.855$ & $1.00$ & $0.0$\\
\\
\hline
\\
 \ [IIA] & $1$ & $3.39$ & $2.45$ & $0.282$ & $0.999$ & $0.0$\\
\\
\ [IIB] & $1$ & $10.7$ & $4.91$ & $0.978$ & $1.00$ & $0.004$\\
\\
\hline
\\
 \ [III] & $0.1$ & $8.66$ & $2.76$ & $0.284$ & $0.985$ & $0.084$\\
\\
\hline
\end{tabular}
\end{center}
\pagebreak

Table-3:Predicted values of $\gamma$ and its corresponding solar and atmospheric mass-squared differences, and mixing
parameters from $m_{LL}$ using the values of $v_{R}$ from Table-1 for case (ii)
\begin{center}
\begin{tabular}{ccccccc}\hline
Type & $\gamma$ & $\bigtriangleup m^{2}_{21}[10^{-5}eV^{2}]$ & 
 $\bigtriangleup m^{2}_{23}[10^{-3}eV^{2}]$ & $\tan^{2}\theta_{12}$
 & $\sin^{2}_2\theta_{23}$ & $\sin\theta_{13}$      \\ \hline
\\
 \ [IA] & $10^{-4}$ & $9.04$ & $2.76$ & $1.012$ & $0.994$ & $0.033$\\
\\
  \ [IB] & $10^{-3}$ & $3.97$ & $2.47$ & $0.268$ & $1.00$ & $0.0$\\
\\
 \ [IC] & $10^{-4}$ & $3.33$ & $2.49$ & $1.760$ & $0.999$ & $0.0$\\
\\
\hline
\\
 \ [IIA] & $1$ & $3.39$ & $2.47$ & $0.289$ & $0.999$ & $0.0$\\
\\
\ [IIB] & $1$ & $9.05$ & $5.04$ & $0.991$ & $0.999$ & $0.002$\\
\\
\hline
\\
 \ [III] & $0.1$ & $5.81$ & $3.02$ & $0.42$ & $0.986$ & $0.078$\\
\\
\hline
\end{tabular}
\end{center}
\pagebreak
Table-4: Predicted values of the $0\nu\beta\beta$ decay mass parameter $|m_{ee}|$
and $\alpha$ for both cases (i) and (ii) from $m_{LL}$ using the values of parameters 
in Table 1-3.

\begin{center}
\begin{tabular}{ccccc}\hline
Type & \multicolumn{2}{c}{Case(i)} & \multicolumn{2}{c}{Case (ii)}\\ 
     & $|m_{ee}|$  & $\alpha$ & $|m_{ee}|$ & $\alpha$             \\\hline
\\
 \ [IA] & $0.084$  & $0.0311$ & $0.084$ & $0.0328$\\
\\
 \ [IB] & $0.3968$  & $0.0160$ & $0.3964$ & $0.016$\\
\\
 \ [IC] & $0.3968$  & $0.0148$ & $0.3968$ & $0.0134$\\
\\
\hline
\\
 \ [IIA] & $0.05$ & $0.0139$ & $0.0501$ & $0.0137$\\
\\
 \ [IIB] & $0.0$ & $0.022$ & $0.0$ & $0.018$\\
\\
\hline
\\
 \ [III] & $0.0$ & $0.0314$  & $0.0$ & $0.0192$\\
\\
\hline
\end{tabular}
\end{center}
\section{Summary and Discussion}
 We summarise  the main points of  this work. We can generate in principle 
the three neutrino mass matrices namely,   
degenerate (Type [I(A,B,C)]), inverted hierarchical 
(Type [II (A,B)]) and normal hierarchical (Type [III]) models, by taking  the diagonal form of the Dirac neutrino mass matrix
and   a non-diagonal
 form of the right-handed Majorana mass matrix in  the canonical seesaw formula (Type I). We then examine whether these good predictions 
are spoiled or not in the presence of the 
left-handed Higgs triplet in Type II seesaw formula; and if so, we find out the least minimum perturbation for retaining 
good predictions which are previously obtained.  We  propose  a kind of  natural selection of the neutrino mass models which have 
``least perturbation'' arising  from Type II seesaw term, in order to retain the good predictions already acquired. Under such hypothesis
 we arrive at the conclusion that 
inverted hierachical model is the most favourable one in nature. Next to it is the normal hierarchy. Degenerate models are badly spoiled 
by the perturbation in Type II seesaw formula, and therefore it is  not favoured  by the natural selection.
Our conclusion also nearly agrees with the calculations using the mass matrices $m_{LR}$ and $M_{RR}$ predicted by  other authors in SO(10) models[34,35]. 
It can be stressed that the method adopted here is also applicable to any neutrino mass matrix obtained using a general non-diagonal texture
 of Dirac mass matrix.

A few comments are in order. Within the Inverted hierarchical model itself, we are having two varieties: Type [IIA] with mass egenvalues $(m_{1}, m_{2},0)$ and 
Type [IIB] with mass eigenvalues $(m_{1},- m_{2},0)$. The present analysis could not distingush which one is more favourable as both predict the same 
$\gamma=1$ which measures the degree of perturbation arising from Type II seesaw formula. This means that for Inverted model, Type I seesaw term donimates over
Type II seesaw term without any fine tuning. 
However within the inverted hierarchy model, Type [IIA] model predicts slightly lesser  solar mixing,  $\tan^{2}\theta_{12}=0.282$ 
 and $\sin\theta_{13}\sim 0.0$ and $|m_{ee}|\sim 0.05$eV 
without any fine tunning.
 Type [IIB] predicts maximal solar mixing  $\tan^{2}\theta_{12}=0.99$,  small CHOOZ angle $\sin\theta_{13}\simeq 0.004$, and $|m_{ee}|\sim 0$.
 This  requires 
some other mechanisms to tone down the solar angle at the cost of increasing $\sin \theta_{13}$ near the  experimental bound. For example, 
taking charged  lepton contribution,  one can have in this case $0.66\geq\tan^{2}\theta_{12}\geq0.49$, corresponding to $0.10\leq|V_{e3}|\leq0.17$ and 
$0.014eV\leq|m_{ee}|\leq0.024eV$ [33].   If we use the lower bound on $|m_{ee}|> 0.013$ eV derived from the SNO data (with salt run)[36], Type [IIB] 
nearly survives. 
Precise measurement of $\sin\theta_{13}$ may help to distinguish these two kinds. This can be tested in the  future long baseline experiments[36].
 As a remark we also point out that unlike Type [IIB] [38], Type [IIA] will be unstable under quantum radiative corrections in MSSM [33]. As emphasised before, 
the present analysis is based on the hypothesis that those models of neutrinos where the  canonical seesaw term is
 dominant over the perturbative term arising from Type II seesaw, are favourable under natural selection. 
The present work is a modest attempt to understand  the correct model of neutrino mass pattern.   

\section*{Acknowledgements}
One of us (N.N.S.) would like to  thank  Prof.Goran Senjanovic for useful discussion and 
Professor Randjbar-Daemi, Head of the  High Energy Group, International Centre for Theoretical Physics, Trieste, Italy,
for kind hospitality at ICTP during my visit under Regular associateship scheme.

\pagebreak
\section*{Appendix A}

We list here for ready reference[16],  the  zeroth-order left-handed Majorana neutrino mass 
matrices with texture zeros, $m_{LL}$, corresponding to three models of neutrinos, viz., 
degenerate (Type [I]), inverted hierarchical (Type [II]) and normal hierarchical (Type [III]). 
These mass matrices are competible with the  LMA MSW solution as well as maximal atmospheric mixings.

\begin{center}
\begin{tabular}{ccc}\hline
Type  & $m_{LL}$ & $m_{LL}^{diag}$\\ \hline \\
\ [IA]    &${ \left(\begin{array}{ccc}
  0 & \frac{1}{\sqrt{2}} & \frac{1}{\sqrt{2}}\\ 
 \frac{1}{\sqrt{2}} & \frac{1}{2} & -\frac{1}{2}\\
 \frac{1}{\sqrt{2}} & -\frac{1}{2} & \frac{1}{2} 
\end{array}\right)}m_{0}$ & $Diag(1,-1,1)m_{0}$\\

\\
\ [IB]    &${ \left(\begin{array}{ccc}
  1 & 0 & 0\\ 
 0 & 1 & 0\\
 0 & 0 & 1 
\end{array}\right)}m_{0}$ & $Diag(1,1,1)m_{0}$\\
\\

\ [IC]    &${ \left(\begin{array}{ccc}
  1 & 0 & 0\\ 
 0 & 0 & 1\\
 0 & 1 & 0 
\end{array}\right)}m_{0}$ & $Diag(1,1,-1)m_{0}$\\ \hline
\\
\ [IIA]    &${ \left(\begin{array}{ccc}
  1 & 0 & 0\\ 
 0 & \frac{1}{2} & \frac{1}{2}\\
 0 & \frac{1}{2} & \frac{1}{2} 
\end{array}\right)}m_{0}$ & $Diag(1,1,0)m_{0}$\\
\\

\ [IIB]    &${ \left(\begin{array}{ccc}
  0 & 1 & 1\\ 
 1 & 0 & 0\\
 1 & 0 & 0 
\end{array}\right)}m_{0}$ & $Diag(1,-1,0)m_{0}$\\ \hline
\\

\ [III]    &${ \left(\begin{array}{ccc}
  0 & 0 & 0\\ 
 0 & \frac{1}{2} & -\frac{1}{2}\\
 0 & -\frac{1}{2} & \frac{1}{2} 
\end{array}\right)}m_{0}$ & $Diag(0,0,1)m_{0}$  \\ 

\\ \hline
\end{tabular}
\end{center}

\pagebreak
\section*{Appendix B}
Here we list the textures of the right-handed neutrino mass matrix $M_{RR}$
along with the left-handed Majorana mass matrix $m_{LL}^{I}$ generated through 
the canonical seesaw formula, Eq.(2), for three different models of neutrinos presented 
in Appendix A. The Dirac neutrino mass matrix is given in Eq.(7) where $m_{f}=m_{\tau}\tan\beta $ for case (i) 
and $m_{f}=m_{t}$ for case (ii), and $m_{0}=m_{f}^{2}/(v_{R})$.  For normal hierarchical model the corresponding matrices are given 
in the main text. These are 
collected from Ref.[27] for ready reference.
\\

\underline{Degenerate model( Type [IA])}:\\

$$M_{RR}= \left(\begin{array}{ccc}
  -2\delta_{2}\lambda^{2m} & ({\frac{1}{\sqrt{2}}}+\delta_{1})\lambda^{m+n} & ({\frac{1}{\sqrt{2}}}+\delta_{1})\lambda^m\\ 
({ \frac{1}{\sqrt{2}}}+\delta_{1})\lambda^{m+n} & ({\frac{1}{2}}+\delta_{1}-\delta_{2})\lambda^{2n} &({ -\frac{1}{2}}+\delta_{1}-\delta_{2})\lambda^n\\
 ({\frac{1}{\sqrt{2}}}+\delta_{1})\lambda^{m} &({ -\frac{1}{2}}+\delta_{1}-\delta_{2})\lambda^n &({ \frac{1}{2}}+\delta_{1}-\delta_{2}) 
\end{array}\right)v_{R}$$
\\
$$-m_{LL}^{I}= \left(\begin{array}{ccc}
  (-2\delta_{1}+2\delta_{2}) & ({\frac{1}{\sqrt{2}}}-\delta_{1}) & ({\frac{1}{\sqrt{2}}}-\delta_{1})\\ 
({ \frac{1}{\sqrt{2}}}-\delta_{1}) & ({\frac{1}{2}}+\delta_{2}) &({ -\frac{1}{2}}+\delta_{2})\\
 ({\frac{1}{\sqrt{2}}}-\delta_{1}) &({ -\frac{1}{2}}+\delta_{2}) &({ \frac{1}{2}}+\delta_{2}) 
\end{array}\right)m_{0}$$
\\
\underline{Degenerate model (Type [IB])}:\\

$$M_{RR}= \left(\begin{array}{ccc}
  (1+2\delta_{1}+2\delta_{2})\lambda^{2m} & \delta_{1}\lambda^{m+n} & \delta_{1}\lambda^{m}\\ 
    \delta_{1}\lambda^{m+n} & (1+\delta_{2})\lambda^{2n} & \delta_{2}\lambda^{n}\\
 \delta_{1}\lambda^{m} & \delta_{2}\lambda^{n} & (1+\delta_{2}) 
\end{array}\right)v_{R}$$
\\
$$-m_{LL}^{I}= \left(\begin{array}{ccc}
  (1-2\delta_{1}-2\delta_{2}) & -\delta_{1} & -\delta_{1}\\ 
    -\delta_{1} & (1-\delta_{2}) & -\delta_{2}\\
 -\delta_{1} & -\delta_{2} & (1-\delta_{2}) 
\end{array}\right)m_{0}$$
\\
\underline{Degenerate model (Type [IC])}:\\

$$M_{RR}= \left(\begin{array}{ccc}
  (1+2\delta_{1}+2\delta_{2})\lambda^{2m} & \delta_{1}\lambda^{m+n} & \delta_{1}\lambda^{m}\\ 
    \delta_{1}\lambda^{m+n} & \delta_{2}\lambda^{2n} & (1+\delta_{2})\lambda^{n}\\
 \delta_{1}\lambda^{m} & (1+\delta_{2})\lambda^{n} & \delta_{2} 
\end{array}\right)v_{R}$$
\\
$$-m_{LL}^{I}= \left(\begin{array}{ccc}
  (1-2\delta_{1}-2\delta_{2}) & -\delta_{1} & -\delta_{1}\\ 
    -\delta_{1} & -\delta_{2} & (1-\delta_{2})\\
 -\delta_{1} & (1-\delta_{2}) & -\delta_{2} 
\end{array}\right)m_{0}$$
\\
\underline{Invereted hierarchical model(Type [IIA])}:\\

$$M_{RR}= \left(\begin{array}{ccc}
  \eta (1+2\epsilon)\lambda^{2m} &\eta\epsilon\lambda^{m+n} & \eta\epsilon\lambda^m\\ 
 \eta\epsilon\lambda^{m+n} & \frac{1}{2}\lambda^{2n} & -(\frac{1}{2}-\eta)\lambda^n\\
 \eta\epsilon\lambda^m & -(\frac{1}{2}-\eta)\lambda^n & \frac{1}{2} 
\end{array}\right)\frac{v_{R}}{\eta}$$
\\

$$-m_{LL}^{I}= \left(\begin{array}{ccc}
  (1-2\epsilon) & -\epsilon & -\epsilon\\ 
 -\epsilon & \frac{1}{2} & (\frac{1}{2}-\epsilon)\\
 -\epsilon & (\frac{1}{2}-\epsilon) & \frac{1}{2} 
\end{array}\right)m_{0}$$


\underline{Inverted hierarchical model(Type  [IIB])}:\\

$$M_{RR}= \left(\begin{array}{ccc}
  \lambda^{2m+7} & \lambda^{m+n+4} & \lambda^{m+4}\\ 
\lambda^{m+n+4} & \lambda^{2n} & -\lambda^n\\
 \lambda^{m+4} & -\lambda^n & 1 
\end{array}\right)v_{R}$$
\\

$$-m_{LL}^{I}= \left(\begin{array}{ccc}
  0 & 1 & 1\\ 
 1 & -(\lambda^3-\lambda^4)/2 & -(\lambda^3+\lambda^4)/2\\
 1 & -(\lambda^3+\lambda^4)/2 & -(\lambda^3-\lambda^4)/2
\end{array}\right)m_{0}$$
\\

The values of the parameters used are: Type IA:  $\delta_{1}=0.0061875$, $\delta_{2}=0.0030625$, $m_{0}=0.4$eV; \ \
Type [IB] and [IC]: $\delta_{1}=3.6\times 10^{-5}$, $\delta_{2}=3.9\times 10^{-3}$, $m_{0}=0.4$eV; \ \
Type [IIA] and [IIB]: $\eta=0.0001$, $\epsilon=0.002$, $m_{0}=0.05$eV. 
\pagebreak


\begin{thebibliography}{99}
\bibitem{1} For a recent review, see  M.Maltoni, T.Schwetz, M.A.Tortola, J.W.F.Valle, {\bf hep-ph/0405172}.
\bibitem{2} A.Yu.Smirnov,  {\bf hep-ph/0402264}.
\bibitem{3} G.Altarelli and F.Feruglio, {\bf hep-ph/0405048}.
\bibitem{4} Super-Kamiokande Collaboration, S.Fukuda  et al.,  Phys. Lett. {\bf B539}, 179 (2002), [{\bf hep-ex/9807003}];\\
            SNO Collaboration,  S.N.Ahmed  et al.,  Phys. Rev. Lett. {\bf 92}, 181301 (2004), [{\bf nucl-ex/0309004}].
\bibitem{5} Super-Kamiokande Collaboration,  Y.Fukuda et al., Phys. Rev. Lett. {\bf 81}, 1562 (1998), [{\bf hep-ex/9807003}].
\bibitem{6} KamLAND Collaboration,  K.Eguchi et al.,  Phys. Rev. Lett. {\bf 90}, 021802 (2003), [{\bf hep-ex/0212021}].
\bibitem{7} CHOOZ Collaboration, M.Apollonio et al., Phys. Lett. {\bf B466}, 415 (1999), [{\bf hep-ex/9907037}].
\bibitem{8} K2K Collaboration, M.H.Ahn et al., Phys. Rev. Lett. {\bf 90}, 041801 (2003), [{\bf hep-ex/0212007}].
\bibitem{9} LSND Collaboration, A.Aguilar et al., Phys. Rev. {\bf D64}, 112007 (2001), [{\bf hep-ex/0104049}].
\bibitem{10} BooNE Collaboration, E.D.Zimmerman, Nucl. Phys. Proc. Suppl. {\bf 123}, 267 (2003), {\bf hep-ex/0211039}.
\bibitem{11} WMAP Collaboration, C.L.Bennett et al., Astrophys. J. Suppl., {\bf 148} (2003) 1;  D.N.Spergel et al., Astrophys. J. Suppl.
            {\bf 148}, 175 (2003); A.Kogut et al, Astrophys. J. Suppl., {\bf 148}, 161 (2003);  G.Hinshaw et al., Astrophys. J. Suppl., {\bf 148}, 135 (2003);
            L.Verde et al., Astrophys. J. Suppl., {\bf 148}, 195  (2003);  H.V.Peiris et al.,  Astrophys. J. Suppl., {\bf 148}, 213  (2003).
\bibitem{12} S.Hannestad, JCAP {\bf 0305}, 004 (2003);  O.Elgaroy and O.Lahav, JCAP {\bf 0304}, 004 (2003);  S.Hannestad, {\bf hep-ph/0310220};
              S.Hannestad, G.Raffelt, {\bf hep-ph/0312154}.
\bibitem{13} The Heidelberg-Moscow Collaboration, H.V.Klapdor-Kleingrothaus et al., Eur. Phys. J, {\bf A12}, 147 (2001); C.E.Aalseth et al., {\bf hep-ex/0202026}.
\bibitem{14} S.M.Bilenky, {\bf  hep-ph/0403245}.
\bibitem{15} H.V.Klapdor-Kleingrothaus et al., Mod. Phys. Lett. {\bf A37}, 2409 (2001); H.V.Klapdor-Kleingrothaus, A.Dietz, I.V.Krivoshena,
              Phys. Lett. {\bf B586}, 198 (2004).
\bibitem{16} G.Altarelli, F.Feruglio, Phys. Rep. {\bf 320}, 295 (1999), {\bf hep-ph/9905536}.
\bibitem{17} M.Gell-Mann, P.Ramond, and R.Slansky, in Supergravity, Proceedings of the Worshop, Stony Brook, New York, 1979, 
             edited by P.van Nieueenhuizen and D.Freedman (North-Holland, Amsterdam, 1979); T.Yanagida, KEK Lectures, 1979 (unpublished); 
             R.N.Mohapatra and G.Senjanovic, Phys. Rev. Lett. {\bf 44}, 912 (1980). 
\bibitem{18} R.N.Mohapatra and  G.Senjanovic, Phys. Rev. {\bf D23}, 165 (1981); 
             G. Lazarides, Q. Shafi, C. Wetterich, Nucl. Phys. {\bf B181}, 287(1981);  C.Watterich, Nucl. Phys. {\bf B187}, 343 (1981).
\bibitem{19} B.Brahmachari and R.N.Mohapatra, Phys. Rev. {\bf D58}, 015001 (1998); Oleg Khasanov and Gilad Perez, Phys. Rev. {\bf D65}, 053007 (2002);
             E.Ma, Phys. Rev. {\bf D69}, 011301 (2004).
\bibitem{20} B.Bajc, Senjanovic and F.Vissani, Phys. Rev. Lett. {\bf 90}, 051802 (2003), {\bf hep-ph/0210207};  H.S.Goh, R.N.Mohapatra, S.P.Ng.,
             Phys.Lett. {\bf B57}, 215 (2003), {\bf hep-ph/0303055}; Phys. Rev. {\bf D68}, 115008 (2003), {\bf hep-ph/0308197}.
\bibitem{21} A.S.Joshipura, E.A.Paschos, W.Rodejohann, JHEP {\bf 0108}, 029 (2001), {\bf hep-ph/0105175};
             Nucl. Phys. {\bf B611}, 227(2001), {\bf hep-ph/0104228}. 
\bibitem{22} R.N.Mohapatra, {\bf hep-ph/0402035};  {\bf hep-ph/0306016}.
\bibitem{23} B.Bajc, G.Senjanovic, F.Vissani, {\bf hep-ph/0402140}.
\bibitem{24} S.Antusch, S.F.King, {\bf hep-ph/0402121}; S. Antusch, S. F. King, {\bf hep-ph/0405093}.
\bibitem{25} W.Rodejohann, {\bf hep-ph/0403236}.
\bibitem{26} T.Hambye, G.Senjanovic, Phys. Lett. {\bf B582}, 73 (2004),  {\bf  hep-ph/0307237}; P.O'Donnell, U.Sarkar, Phys. Rev. {\bf D49}, 2118 (1994).
\bibitem{27} N.Nimai Singh and M.Patgiri, IJMP {\bf A17}, 3629 (2002); M.Patgiri and N.Nimai Singh, IJMP {\bf A18}, 443 (2003).
\bibitem{28} K.S.Babu, B.Dutta, R.N.Mohapatra, Phys. Rev. {\bf D67}, 076006 (2003), {\bf hep-ph/0211068}.
\bibitem{29} For a discussion, see, I.Dorsner, S.M.Barr, Nucl. Phys. {\bf B617}, 493 (2001); S.M.Barr, I.Dorsner, Nucl. Phys. {\bf B585}, 79 (2000).
\bibitem{30} E.Kh.Akhmedov, M.Frigerio, A.Yu.Smirnov,  JHEP {\bf 0309}, 021(2003), {\bf hep-ph/0305322}.
\bibitem{31} D.Falcone, Phys. Lett. B479, 1 (2000), hep-ph/0204335.
\bibitem{32} H.Murayama,{\bf  hep-ph/02010022}; A.de Gouvea, A.Friedland, H.Murayama, Phys. Lett. {\bf B490}, 125 (2000).
\bibitem{33} M.Patgiri and N.Nimai Singh, Phys. Lett. {\bf B567}, 69 (2003).
\bibitem{34} Carl H.Albright and S.M.Barr, {\bf hep-ph/0404095}.
\bibitem{35} K.S.Babu, S.M.Barr, Phys.Rev.Lett.{\bf 85}, 1170 (2000).
\bibitem{36} H.Murayama and C.Pena-Garay, {\bf hep-ph/0309114}.
\bibitem{37} An incomplete list, Ernest Ma, Phys.Rev.{\bf D69},011301(2004),{\bf hep-ph/0308092}; Utpal Sarkar, {\bf hep-ph/0403276}; 
M.K.Parida, B.Purkayastha, C.R.Das,
             B.D.Cajee,{\bf hep-ph/0210270}; B.Bajc, G.Senjanovic, F.Vissani, Phys.Rev.Lett.{\bf 90},051802(2003);
 A Melfo, G.Senjanovic, Phys.Rev.{\bf D68},03501(2003).
\bibitem{38} S. F. King, N. Nimai Singh, Nucl. Phys. {\bf B596}, 81(2001).
\bibitem{39} S. F. King, N. Nimai Singh, Nucl. Phys. {\bf B591}, 3(2000).
\end{thebibliography}
\end{document}